# Quasi-1D Electronic Metadevices with Enhanced Electrical Properties

Abdallah Abushawish, Ziwen Huang, and Mohammad Samizadeh Nikoo

**Electronic metadevices leveraging sub-wavelength metallic features have shown great potential for high-frequency switching. Theoretical analysis based on a one-dimensional (1D) model indicates that reducing the size of subwavelength features can enhance electrical properties, such as contact resistance and switching cutoff frequency. Here, we report higher-than-expected contact resistance in electronic metadevices when the subwavelength feature size is aggressively downscaled. We attribute this effect to transverse currents in the two-dimensional electron gas (2DEG) running parallel to the stripe width. We present the first experimental demonstration of a metadevice governed by a one-dimensional model, which we refer to as a quasi-1D electronic metadevice and show that it enables enhanced electrical performance. Our findings pave the way to design next generation electronic metadevice switches with applications in future telecommunication circuits.**

Next-generation wireless communication systems demand high-performance electronic switches to support increasing data rates, ultra-low latency, and expanded frequency coverage [1]-[4]. Switches are fundamental to telecommunication circuits, influencing the performance of wireless networks, including 5G, 6G, and beyond, where ultrafast operation is essential [5]-[7]. However, as operating frequencies continue to rise, conventional switches face inherent limitations [8], [9]. One major challenge is the contact resistance at metal-semiconductor interfaces. This resistance together with parasitic capacitances limit the cutoff frequency figure-of-merit (FOM) ($f_C$) of semiconductor switches [6][9]-[13].

Electronic metadevices leveraging quasi-electrostatic subwavelength control of electric fields and currents offer an alternative approach for efficiently coupling metals to semiconductors [14]-[17]. This concept relies on patterning metallic electrodes to create subwavelength features, which enables ultra-low contact resistances down to 10 Ω μm at terahertz frequencies [14]. It was shown that shrinking the subwavelength features size (*L*) in electronic metadevice (Fig. 1a) can further enhance key performance metrics such as contact resistance and cutoff frequency FOM [16].

Here, we report a higher-than-expected contact resistance when aggressively scaling down the subwavelength feature of electronic metadevices. We attribute this to unintended current flow parallel to the stripes (transverse current, $J_y$ in Fig. 1b), which can lead to a contact resistance more than 10 times higher than that predicted by the 1D theory when the subwavelength feature is scaled down to 200 nm. While the theoretical predictions derived from a 1D model suggest that reducing these dimensions should lower contact resistance [16], our findings reveal that two-dimensional (2D) current flow introduces additional dissipation.

We propose quasi-1D electronic metadevices, where the 2D electron gas (2DEG) is confined strictly beneath subwavelength-scale stripes. This structural change alters the current flow, significantly enhancing the ON-state conductance. We have enhanced the switching cutoff frequency of conventional electronic metadevices (Figs. 2a-b) by a factor of three through the implementation of quasi-1D electronic metadevice concepts (Fig. 2e). Our findings lay the groundwork for designing the next generation of electronic metadevices. Moreover, while electronic metadevices have been known for their excellent performance at millimeter-wave and terahertz frequencies, our work establishes a new paradigm for high-performance microwave metadevices, which are valuable for telecommunication networks operating at sub-100 GHz.

Fig. 2a shows a top-view schematic of a conventional electronic metadevice with a stripe length of $L$ = 1 μm. Fig. 2b shows a finite element method (FEM) simulation of the current density in the 2DEG, obtained from a three-dimensional frequency-domain solution of the electrostatic equations using COMSOL. The results display the longitudinal ($J_x$, left) and transverse ($J_y$, right) components. We used frequency-domain simulations in COMSOL to solve electrostatic equations in 3D. The transverse component is insignificant compared to the longitudinal component, except at the end of the stripes. Fig. 2c presents the schematic of an electronic metadevices with reduced stripe length of $L$ = 0.2 μm. The corresponding current densities shown in Fig. 2d reveal a substantial increase in the transverse component across the device layout. This current does not contribute to the conduction between the two terminals but results in significant ohmic losses in the 2DEG.

Fig. 2e shows a top-view schematic of an electronic metadevice similar to that in Fig. 2c, where the 2DEG exists solely beneath the stripes. Fig. 2f presents the resulting current densities, where transverse current is completely dominant. We refer to this structure as a quasi-1D electronic metadevice, as it represents the experimental demonstration of the 1D collective effect theoretically demonstrated in [16].

Fig. 2g (solid points) highlights the impact of additional ohmic losses caused by transverse currents in conventional electronic metadevices. In this case, the normalize contact resistance ($R_C$) initially decreases by shrinking $L$, however, it sharply increases for $L$ < 1 μm, deviating from the theoretical results based on the 1D model [16]. These results show that the effect of transverse currents can hinder the superior performance of electronic metadevices. The quasi-1D design, however, maintains contact resistance values that closely align with theoretical expectations of the 1D model, significantly outperforming conventional electronic metadevices. We note that both conventional and quasi-1D electronic metadevices operate under Maxwell's equations in the electrostatic regime. However, the key distinction lies in the enforced lateral symmetry of the quasi-1D structure, which ensures $\partial/\partial y$ = 0. This reduction allows the device to follow the one-dimensional model developed in our earlier work [16].

Electronic metadevices were fabricated on an InAlN/GaN heterostructure grown on a SiC substrate. The epitaxy features a 5-nm-thin InAlN barrier layer, sandwiched between a 1-nm GaN cap and a 1-nm AlN spacer, from top and bottom, respectively. A chlorine-based inductively coupled plasma (ICP) etching step ($BCl_3/Cl_2$ at 20/10 sccm) at RF power of 100 W defined the mesa region after photolithography. Electron beam lithography (EBL) was then used to define the pads and stripe regions. Finally, Nickel deposition followed by lift-off formed the Schottky contacts for both terminals.

Fig. 3a presents an overview of the measurement setup, featuring an optical microscope image of a quasi-1D metadevice connected to ground-signal-ground (GSG) probes. Fig. 3b presents the measured ON-resistance ($R_{ON}$) as a function of frequency for both conventional and quasi-1D electronic metadevices, both with identical stripe length of $L$ = 0.3 μm. The quasi-1D metadevice demonstrates a markedly lower $R_{ON}$, which is almost independent of frequency. Notably, this performance improvement is achieved despite the quasi-1D metadevice having a smaller effective width ($W_{eff} \cong$ 33 μm) compared to the conventional metadevice ($W_{eff} \cong$ 41 μm).

Fig. 4a presents the measured normalized $R_{ON}$ and OFF-state capacitance ($C_{OFF}$) for conventional and quasi-1D electronic metadevices as a function of stripe length. Lower $R_{ON}$ and $C_{OFF}$ are beneficial for radiofrequency switches to achieve a higher cutoff frequency ($f_C$ = 1/(2π$R_{ON}C_{OFF}$)). Quasi-1D electronic metadevices show lower $R_{ON}$ across the entire range of stripe lengths. However, the superiority of the quasi-1D design is more pronounced for smaller stripe lengths ($L \ll$ 1 μm). There is a slight increase in the $R_{ON}$ of quasi-1D devices when $L$ approaches 0.3 μm, which is attributed to the series resistance of narrow metallic stripes ($W$ = 3 $\mu m \gg L$).

Fig. 4b shows the ratio of transverse-to-longitudinal current density intensities, defined as $\int_A J_y^2 \, dx \, dy / \int_A J_x^2 \, dx \, dy$), as a function of the stripe length. The results indicate that the relative contribution of transverse currents increases as the stripe length decreases in conventional metadevices, whereas it remains negligible in quasi-1D electronic metadevices.

We note that while normalized $C_{OFF}$ is slightly higher in the quasi-1D metadevice for $L < 1$ μm, the substantial reduction in $R_{ON}$ results in a superior cutoff frequency as shown in Fig. 4c, from 1.7 THz to 5.6 THz. Fig. 5a benchmarks the quasi-1D metadevice against the conventional metadevice and mainstream electronic devices on a $C_{OFF}$-$R_{ON}$ plane, indicating superior FOM for quasi-1D electronic metadevices.

We demonstrated quasi-1D electronic metadevices with superior electrical performance by effectively suppressing transverse currents running parallel to the stripe width. An earlier theoretical study [16] predicted the possibility of realizing a meta effect in a one-dimensional structure without transverse currents. Quasi-1D electronic metadevices represent the first experimental validation of this concept. We showed that the quasi-1D structure significantly reduces $R_{ON}$, which increases the switching cutoff frequency FOM ($f_C$). Quasi-1D electronic metadevices achieve a cutoff frequency up to three times higher than that of the conventional metadevice. These findings offer key insights into the design of electronic metadevices with enhanced electrical properties, particularly for operation below 100 GHz.

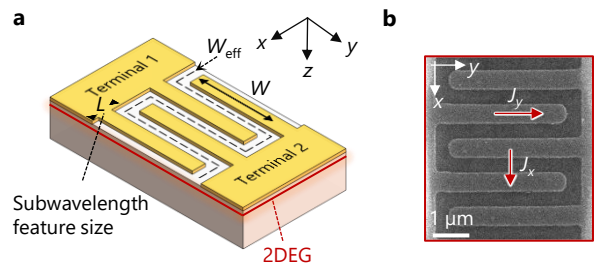

Fig. 1. (a) 3D schematic of a conventional electronic metadevice. (b) Scanning Electron Microscopy (SEM) image of an electronic metadevice indicating the longitudinal ($J_x$) and transverse ($J_y$) currents.

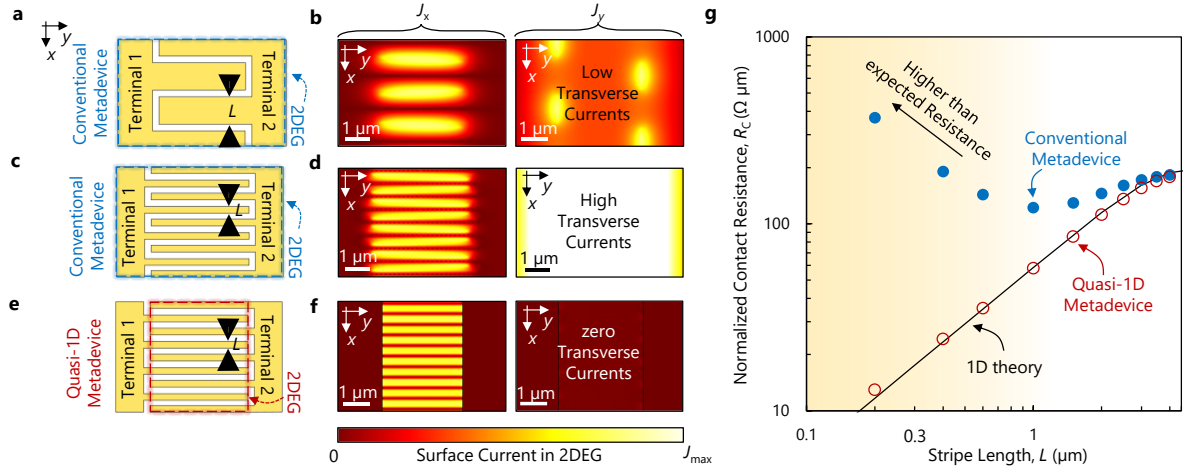

Fig. 2. (a) Schematic of a conventional electronic metadevice ($L$ = 1 µm) with 2DEG extending across the entire device. (b) The corresponding longitudinal (left) and transverse (right) current densities in the 2DEG. (c) Schematic of a conventional electronic metadevice with $L$ = 0.2 µm. (d) The corresponding longitudinal (left) and transverse (right) current densities in the 2DEG. (e) Schematic of a quasi-1D metadevice ($L$ = 0.2 µm) with 2DEG confined under the stripes. (f) The corresponding longitudinal (left) and transverse (right) current densities in 2DEG. (g) Normalized contact resistance versus stripe length for conventional and quasi-1D electronic metadevices. Simulation: discrete points, theory: solid line ($R_{sh}$ = 200 Ω/□).

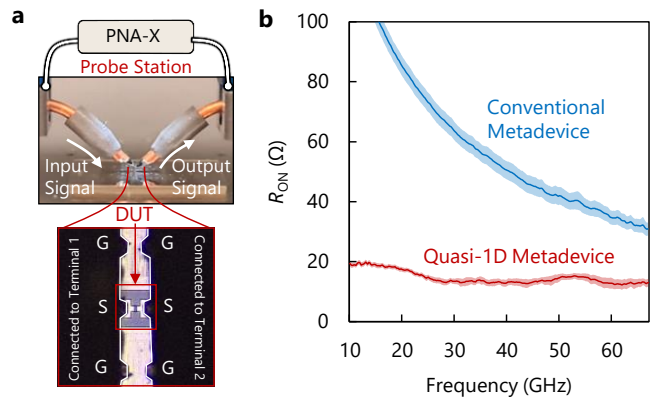

Fig. 3. (a) Schematic of the measurement setup to measure scattering (S)-parameters of electronic metadevices (top panel). Optical image of an electronic metadevice connected to GSG probes (bottom panel). (b) Extracted $R_{ON}$ over frequency for conventional and quasi-1D electronic metadevices. Shaded regions represent $\pm\sigma$ (standard deviation) from the average $R_{ON}$, calculated across 6 devices.

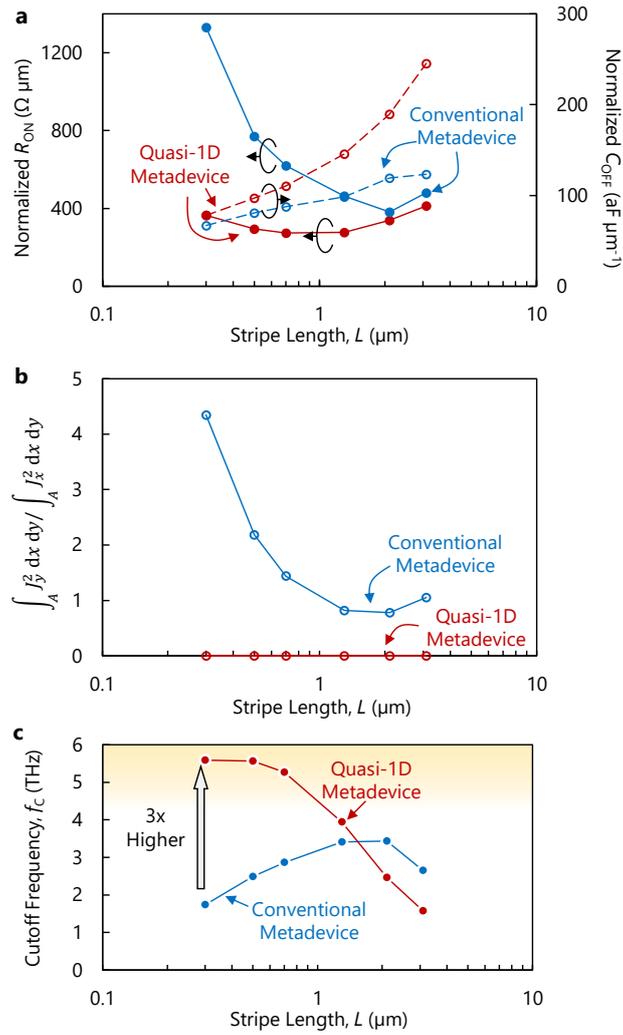

Fig. 4. (a) Normalized $R_{ON}$ (left vertical axis) and $C_{OFF}$ (right vertical axis) for conventional and quasi-1D electronic metadevices as a function of stripe length, measured at 67 GHz. (b) Ratio of transverse-to-longitudinal current intensities as a function of stripe length. $A$ denotes the area of the 2DEG. (c) $f_C$ for conventional and quasi-1D electronic metadevices as a function of stripe length.

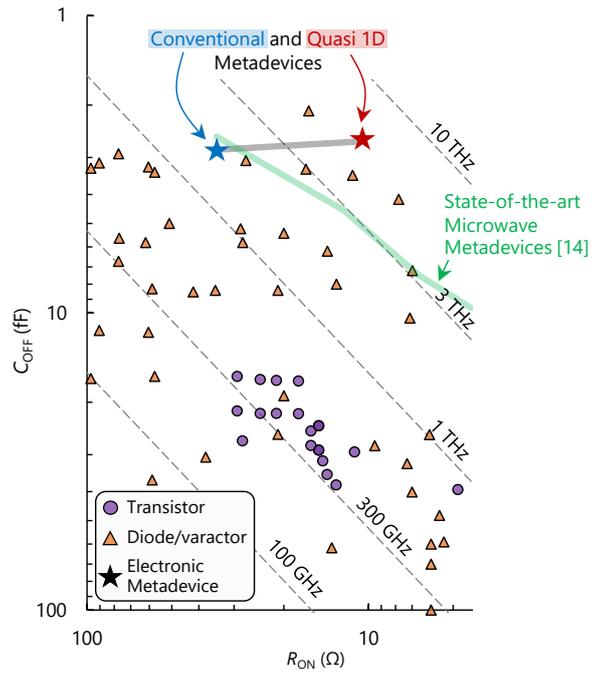

Fig. 5. Benchmark of the quasi-1D electronic metadevice against the conventional metadevice and mainstream electronic devices.